\documentclass[12pt,preprint]{aastex}

\def\vec#1{\mbox{\boldmath $#1$}}

\slugcomment{To appear in the Astrophysical Journal}

\shorttitle{NLFF Extrapolation of the Coronal Magnetic Field}
\shortauthors{Inoue et al.}

\begin{document}


\title{Nonlinear Force-Free Extrapolation of the Coronal Magnetic Field \\
       Based on the MHD Relaxation Method}


\author{S.\ Inoue}
\affil{ School of Space Research, Kyung Hee University, Yongin, Gyeonggi-do 
        446-701,\\ Republic of Korea}
\email{inosato@khu.ac.kr}

\author{T.\ Magara}
\affil{ School of Space Research, Kyung Hee University, Yongin, Gyeonggi-do
        446-701,\\ Republic of Korea}

\author{V.\ S.\ Pandey}
\affil{ Department of Physics, National Institute of Technology, Dwarka, 
        Sector-9, Delhi-110077, India}

\author{D.\ Shiota}
\affil{Solar-Terrestrial Environment Laboratory, Furo-Cho, Chikusa-ku Nagoya
       464-8601, Japan}
\altaffiltext{1}{ Computational Astrophysics Laboratory, RIKEN(Institute
       of Physics and Chemical Research),\\ Wako, Saitama 351-0198, Japan}
\altaffilmark{1}

\author{K.\ Kusano}
\affil{Solar-Terrestrial Environment Laboratory, Furo-Cho, Chikusa-ku Nagoya
       464-8601, Japan}
\altaffiltext{2}{Japan Agency for Marine-Earth Science and Technology
                (JAMSTEC),Yokohama, Kanagawa 236-0001, Japan}
\altaffilmark{2}

\author{G.\ S.\ Choe, and K.\ S.\ Kim}
\affil{ School of Space Research, Kyung Hee University, Yongin, Gyeonggi-do 
        446-701,\\ Republic of Korea}

  \begin{abstract}
   We develop a nonlinear force-free field (NLFFF) extrapolation code 
  based on the magnetohydrodynamic (MHD) relaxation method. We extend the 
  classical  MHD relaxation method in two important ways. First, we 
  introduce an algorithm initially proposed by \cite{2002JCoPh.175..645D} 
  to effectively clean the numerical errors associated with 
  $\nabla\cdot\vec{B}$. Second, the multi-grid type method is implemented in 
  our NLFFF to perform direct analysis of the high-resolution magnetogram 
  data. As a result of these two implementations, we successfully 
  extrapolated the high resolution force-free field introduced by 
  \cite{1990ApJ...352..343L} with better accuracy in a drastically shorter 
  time. We also applied our extrapolation method to the MHD solution obtained 
  from the flux-emergence simulation by \cite{2012ApJ...748...53M}. We found 
  that NLFFF extrapolation may be less effective for reproducing areas higher 
  than a half-domain, where some magnetic loops are found in a state of 
  continuous upward expansion. However, an inverse S shaped structure 
  consisting of the sheared and twisted loops formed in the lower region can 
  be captured well through our NLFFF extrapolation method. We further discuss 
  how well these sheared and twisted fields are reconstructed by estimating 
  the magnetic topology and twist quantitatively.
  \end{abstract}

  \section{Introduction}
  Solar active phenomena such as solar flares, coronal mass ejections (CMEs), 
  and filament eruptions are widely attributed to the release of magnetic 
  energy in the solar corona ( \citealt{2002A&ARv..10..313P} and 
  \citealt{2011LRSP....8....6S}). Many theoretical and numerical models have 
  been proposed for understanding their dynamics and triggering mechanisms 
  (The details are summarized in 
   \citealt{2009JGRA..11400B09L};
   \citealt{2011LRSP....8....1C};  
   \citealt{2011LRSP....8....6S}.). 
  However, a number of issues remain unanswered. For instance, we still 
  do not have a proper view of the three-dimensional (3D) coronal magnetic 
  field related to an active region. Earlier studies based on analytical 
  or numerical models were categorized in terms of the 'loss of stability' 
  or 'loss of equilibrium', however, there are also a few cases that strongly 
  depend on the magnetic configurations because the coronal magnetic field 
  deduced from observational images is very complicated, and one cannot 
  extract its physical essence in term of the simplified analytical models. 
  Therefore,it becomes important to construct the coronal magnetic field on 
  the basis of a numerical model by using the observational data and to 
  investigate the physical condition of the equilibrium state before the 
  flare. 

  Unfortunately, the 3D coronal magnetic field cannot be directly observed 
  even with the state-of-art solar physics satellite, whose observations 
  can currently provide only the vector field on the photosphere. For 
  these reasons, force-free extrapolation has been performed based on a 
  vector field. The force-free field is expressed as follows;
  \begin{equation}
  \vec{\nabla}\times \vec{B} = \alpha (r) \vec{B},  
  \label{ff_eq_1}
  \end{equation}
    or
  \begin{equation}
  \vec{B} \cdot \vec{\nabla} \alpha (r)  = 0,  
  \label{ff_eq_2}
  \end{equation}
  and it has been widely accepted as an approximation of the coronal magnetic 
  field because the value of the plasma $\beta$ is very low 
  ($10^{-1} \sim 10^{-2}$) in the solar corona. The force-free state is 
  classified into three energy levels. One of them, called the potential 
  field is the current free state {\it i.e.} $\alpha(r)=0$, hence, this
  corresponds to the minimum energy state. The linear force-free field (LFFF) 
  has a uniform distribution of $\alpha(r)$, and this energy level is higher 
  than the potential field. However the observed $\alpha$ is generally a  
  function of space on the photosphere.

   Observations at various wavelengths often reveal a localized strong shear 
  field close to the neutral line in the active regions before a flare
  ({\it e.g.,} \cite{1984SoPh...91..115H};\cite{2007PASJ...59S.785S}). From 
  these observational images and results, we know that the potential and 
  LFFF cannot adequately explain the coronal magnetic field before the 
  flare; thus, the nonlinear force-free field (NLFFF) has been considered to 
  model the active region's magnetic field. Because the force-free equation 
  is essentially nonlinear, it is not straightforward to solve it for the 
  coronal magnetic field. The solution is obtained only numerically through 
  an iteration process for a fixed vector field on the bottom boundary, 
  whereas the potential field and LFFF are calculated easily from the normal 
  component of the magnetic field on the solar surface
  (\citealt{1989SSRv...51...11S}). Various methods of obtaining NLFFF 
  solutions have been proposed and developed. For the sake of brevity, we do 
  not review them thoroughly here; interested readers are urged to see the 
  comprehensive reviews on this topic by 
  \cite{2006SoPh..235..161S}, \cite{2008SoPh..247..269M} or 
  \cite{2012LRSP....9....5W}.

   Representative methods for extrapolating the force-free coronal magnetic 
   field include time-evolutionary methods as well as iterative methods 
  ( {\it e.g.,} the boundary integral method of  
    \citealt{1997SoPh..174...65Y}; 
    \citealt{2000SoPh..195...89Y}
   and the Grad-Rubin methods of 
    \citealt{1981SoPh...69..343S};
    \citealt{1997SoPh..174..129A};
    \citealt{2004SoPh..222..247W};
    \citealt{2006A&A...446..691A}),
   which iterate an equation to find a solution(\ref{ff_eq_1}).
   \cite{1994ApJ...422..899M} and \cite{1994ASPC...68..225M} 
   developed the magnetohydrodynamic (MHD) relaxation method, which directly 
   solves the MHD equations under the zero $\beta$ approximation 
   (\citealt{1988ApJ...328..830M}). These equations include the resistivity 
   which allows the field lines to change their topology rapidly toward a 
   force-free state. The calculation begins with the construction of a 
   potential field from the normal component of the magnetic field on the 
   photosphere and a force-free state is obtained by controlling the 
   transverse electric field and keeping the  magnetic flux $B_z$ according 
   to an induction equation toward the normal component of the current 
   density deduced from the vector field. 
   \cite{2011ApJ...727..101J} and \cite{2012ApJ...749..135J} recently 
   developed a force-free extrapolation code based on the MHD relaxation 
   method that includes the gas pressure, viscous and resistive terms.
   That code is implemented into the space-time conservation-element and 
   solution-element method constructed using full MHD system and a modern 
   high-performance numerical method (\citealt{2007ApJ...655.1110F}; 
   \citealt{2010ApJ...723..300F}). 
   
  \citealt{1996ApJ...473.1095R} developed a force-free extrapolation 
  code that included an induction equation with a hypothetical velocity 
  against the Lorentz force, which is a simplified formation from 
  \cite{1994ApJ...422..899M} and \cite{1994ASPC...68..225M}. This formula 
  was originally introduced by \cite{1986ApJ...309..383Y} to obtain a 
  magneto-frictional method. In addition, this calculation is classified into 
  two phases, stress and relaxation after a potential field is constructed as 
  an initial state from the normal component of the magnetic field on the 
  photosphere. In the stress phase a Lorentz force is injected from the 
  bottom boundary so that the transverse components of the vector potential 
  approach the observed transverse field. In the relaxation phase, the upper 
  coronal field relaxes toward a force-free state under the fixed bottom 
  boundary. Because of these two combined effects in this method, it is called 
  stress and relaxation method. Some authors have already implemented this  
  method into their own code 
  (\citealt{2005A&A...433..335V};  \citealt{2011ApJ...727..101J}).    
  The resistivity included in the induction equation permits the magnetic 
  reconnection to accelerate the process of the force-free state; thus it 
  plays the  same role as in 
  \cite{1994ApJ...422..899M} and \cite{1994ASPC...68..225M}. 
  \cite{2005A&A...433..335V} introduced the magnetic induction field 
  vector to replace the vector potential. This make it easier to implement 
  the boundary condition than in the original stress and relaxation method 
  of \cite{1996ApJ...473.1095R}. They applied their extrapolation 
  method to the twisted loops obtained from \cite{2003A&A...406.1043T}   
  and found that the NLFFF performs reasonably well for reconstructing of 
  the twisted loops in the localized area close to the neutral line. The 
  improved code of the \cite{2007SoPh..245..263V} and 
  \cite{2010A&A...519A..44V} is applied to the ideal force-free field 
  introduced by \cite{1990ApJ...352..343L} and also to a more complex 
  situation by \cite{1999A&A...351..707T}.
  \cite{2004ApJ...612..519V} inserts a twisted magnetic flux tube into a 
  potential field and the magnetofriction (\citealt{2000ApJ...539..983V}) 
  drives as system toward a force-free state without the transverse component 
  on the photosphere; its magnetic configuration is then compared with  
  observational images. 
  Another method is an optimization method originally proposed and developed 
  by \cite{2000ApJ...540.1150W}  and an improved version of it presented by 
  \cite{2004SoPh..219...87W}, which minimizes a function consisting of 
  divergence-free and force-free fields. Although the basic equation in the  
  optimization method also includes a higher-oder differential equation, 
  which is difficult to solve even numerically, highly accurate reconstruction
   is recorded in some papers ({\it e.g,}  \citealt{2006SoPh..235..161S} )    
                
    Recently the Solar Optical Telescope(SOT) on board {\it Hinode} 
   (\citealt{2007SoPh..243....3K} and \citealt{2008SoPh..249..167T}) can 
   provide images of the vector field with a high spatial resolution (more 
   than 1K pixels). Moreover, the Helioseismic and Magnetic Imager (HMI) on 
   board the {\it Solar Dynamics Observatory (SDO)} can provide vector field 
   data with a high temporal resolution (every 12s), which enables analysis 
   of the NLFFF in unprecedented temporal resolution. Thus it would be 
   interesting to see the performance of the NLFFF with these high-resolution 
   data. 

   The purpose of this study is to develop an extrapolation code for the NLFFF
   that accelerates the calculation time even when these high-resolution data 
   are used. We extended the original MHD relaxation method of  
   \cite{1994ApJ...422..899M} and \cite{1994ASPC...68..225M} in two important 
   ways. First, we implemented an algorithm to prevent the deviation from
   $\vec{\nabla} \cdot \vec{B}=0$ introduced by \cite{2002JCoPh.175..645D}, 
   in which the time dependent term corresponding to $\phi$ (see equation 
   \ref{div_eq}) is used to remove the numerical error of 
   $\vec{\nabla} \cdot \vec{B}$. Second, we implemented a multi-grid-type 
   method (\citealt{1977_Brandt}) to more rapidly propagate information on 
   the boundary condition at a larger scale inside the domain than in the 
   smaller component, which accelerates the speed towards a force-free state. 
   The accuracy and reliability are investigated by using the ideal force-free 
   solution introduced by \cite{1990ApJ...352..343L}. We further apply 
   our extrapolation code to the MHD solution obtained from the flux emergence
   simulation by \cite{2012ApJ...748...53M} to investigate the reliability 
   of the NLFFF extrapolation in a real physical situation. 

   This paper is constructed as follows. The extrapolation and numerical 
   methods are described in Section2. The result of the reconstruction 
   using the Low $\&$ Lou solution is presented in Section 3 and whereas the
   MHD solution from \cite{2012ApJ...748...53M} is shown in Section 4. 
   Finally, some important discussions and conclusions are summarized in 
   Section 5.
 
  \section{Numerical Method}
   We developed an NLFFF extrapolation code based on  MHD relaxation 
  by implementing the multi-grid-type procedure and an algorithm for cleaning 
  the errors related to $\vec{\nabla}\cdot \vec{B}$. We demonstrated  the 
  performance of this method in our previous studies, {\it e.g.}, 
  \cite{2011ApJ...738..161I}, \cite{2012ApJ...747...65I}, 
  \cite{2012ApJ...760...17I} and \cite{2013arXiv1304.8073I}. Nevertheless, 
  several issues were not covered extensively in the previous works and 
  require more detailed explanations.

  This method is formulated using the zero-beta MHD equations where the gas 
  pressure and gravity are neglected (\citealt{1988ApJ...328..830M}) to 
  achieve a force-free state. In this study, we numerically solve the 
  following equations
  \begin{equation}
  \frac{\partial \vec{v}}{\partial t} 
                     = - (\vec{v}\cdot\vec{\nabla})\vec{v}
                       +  \frac{1}{\rho}\vec{J}\times\vec{B}
                        + \nu\vec{\nabla}^{2}\vec{v}.
  \end{equation}

  \begin{equation}
  \frac{\partial \vec{B}}{\partial t} 
                        =  \vec{\nabla}\times(\vec{v}\times\vec{B}
                        -  \eta\vec{J})
                        -  \vec{\nabla}\phi, 
  \label{induc_eq}
  \end{equation}

  \begin{equation}
  \vec{J} = \vec{\nabla}\times\vec{B}.
  \end{equation}
  
  \begin{equation}
  \frac{\partial \phi}{\partial t} + c^2_{h}\vec{\nabla}\cdot\vec{B} 
    = -\frac{c^2_{h}}{c^2_{p}}\phi,
  \label{div_eq}
  \end{equation}
  where $\vec{B}$ is the magnetic flux density, $\vec{v}$ is the velocity, 
  $\vec{J}$ is the electric current density, $\rho$ is the pseudo density, 
  and $\phi$ is the convenient potential. The pseudo density is assumed to 
  be proportional to $|\vec{B}|$ in order to ease the relaxation by equalizing
  the Alfven speed in space. The last equation (\ref{div_eq}) introduced 
  by \cite{2002JCoPh.175..645D} plays a crucial role in avoiding deviation 
  from $\nabla\cdot\vec{B}=0$. From equations (\ref{induc_eq}) and 
  (\ref{div_eq}), we can obtain the following equation: 
  \begin{equation}
  \frac{\partial^2 (\vec{\nabla}\cdot\vec{B})}{\partial t^2}
  + \frac{c_h^2}{c_p^2}\frac{\partial (\vec{\nabla}\cdot\vec{B})}{\partial t}
  = c_h^2\vec{\nabla^2}(\vec{\nabla}\cdot\vec{B}),
  \end{equation}
  which illustrates the propagating and diffusing nature of the numerical 
  errors related to $\vec{\nabla}\cdot\vec{B}$, where $c_h$ and $c_p$  
  correspond to the advection and diffusion coefficients, respectively. 
  The main advantages of this method are that (I) it can be very easily 
  implemented in our numerical code without the need for many improvements 
  and (II) it accelerates the process of removing errors and does not take as  
  much time as that required to remove errors by solving the Poisson equation 
  (see \citealt{2000JCoPh.161..605T} or  \citealt{1995JGR...10012057T}).

   The length, magnetic field, velocity, time and electric current density 
  are normalized by   
  $L_0$ ,  
  $B_0$ , 
  $V_{A}\equiv B_{0}/(\mu_{0}\rho_{0})^{1/2}$,    
  $\tau_{A}\equiv L_{0}/V_{A}$, and     
  $J_0=B_{0}/\mu_{0} L_{0}$,  
  respectively.  The non-dimensional viscosity $\nu$ is set to a constant,   
  $(1.0\times 10^{-3})$, and the non-dimensional resistivity $\eta$ is 
  given by the functional 

  \begin{equation}
  \eta = \eta_0 + \eta_1 \frac{|\vec{J}\times\vec{B}||\vec{v}|^2}{\vec{|B|^2}},
  \end{equation} 
  where, $\eta_0$ depends on each case, as shown in table \ref{tbl-1}, and 
  $\eta_1$ is fixed at $1.0\times 10^{-3}$ in non-dimensional units. The 
  second term is introduced to accelerate the relaxation to the force-free 
  state particularly in the weak field region. The parameters $c_p^2$ are 
  fixed at constants 0.1, whereas $c_h^2$ varies according to table 
  \ref{tbl-1}.
    
  The velocity field is adjusted in such a way that it does not correspond 
  to a large value; otherwise, it would affect the Courant-Friedrichs-Lewy
  condition. We define  
  $v^{*} = |\vec{v}|/|\vec{v}_{A}|$ and if the value of $v^{*}$ becomes 
  larger than the value of $v_{max}$ as given in table \ref{tbl-1}, the 
  velocity is modified as follows:
  \begin{equation}
  \vec{v} \Rightarrow \frac{v_{max}}{v^{*}} \vec{v}.
  \end{equation}

   Two different types of boundary conditions are applied in this 
  study to extrapolate the 3D coronal magnetic field. The first is that 
  all six boundaries are set to the exact solutions obtained from 
  Low $\&$ Lou. We denote this boundary condition as EX. In the second, 
  only the bottom boundary is set to the exact solution from Low $\&$ Lou 
  or \cite{2012ApJ...748...53M} and the other boundaries are assumed to act
  like rigid walls; {\it i.e.,} the normal component of the magnetic field 
  is fixed at the original solutions, and the tangential component is 
  determined by the induction equation as described in equation 
  (\ref{induc_eq}). We denote this boundary condition as RW, which is less 
  information than it in EX. In all cases, the velocity field ($\vec{v}$) 
  is set to zero on all the boundaries. A Neumann-type boundary condition 
  ($\partial_n \phi = 0$) is applied for the potential $\phi$ at all the 
  boundaries, where $\partial_n$ represents the derivative for the normal 
  direction on the surface. The initial condition is given by a potential 
  field calculated from the normal component on all the boundaries for all 
  cases. 

   In the ideal force-free cases (the Low $\&$ Lou solution), we apply the 
  exact solutions directly on each boundary surface. On the other hand, in 
  the MHD solution obtained from \cite{2012ApJ...748...53M}, the handling of 
  the bottom boundaries differs from that in the ideal force-free case except 
  for the normal component. In this case, we introduce a procedure analogous 
  to the stress and relaxation method. The transverse component 
  $\vec{B}_{BC}$ is defined as a linear combination of $\vec{B}_{obs}$ and 
  $\vec{B}_{pot}$ on the bottom surface as follows:  
  \begin{equation}
  \vec{B}_{BC} = \gamma \vec{B}_{obs} + (1-\gamma) \vec{B}_{pot},
  \label{step_eq}
  \end{equation}
  where  $\vec{B}_{obs}$ and $\vec{B}_{pot}$ are the transverse  components 
  of the observational (MHD solution in this study) and the potential field, 
  respectively. $\gamma$ is a coefficient ranging from rage of 0 to 1. When 
  $R = \int |\vec{J}\times\vec{B}|^2 dV$, which is introduced as an indication
  for the force-free state, drops below a critical value denoted by 
  $R_{min}$ during an iteration, then $\gamma$ grows according to 
  $\gamma = \gamma + d\gamma$, where d$\gamma$ is also given as a parameter. 
  $\gamma$ becomes equal to 1; then $\vec{B}_{BC}$ can be completely 
  consistent with the observational data.
 
   As for the numerical method, the spatial derivative is approximated by 
  the second-order finite difference and a time integration is conducted,
  using the Runge-Kutta-Gill method to fourth-order accuracy. Furthermore, 
  we adapt the multi-grid-type method to accelerate the procedure for  
  achieving a force-free state. This method contains the several distinct 
  numerical grids with different resolutions; the first calculation starts 
  using the coarsest one to obtain a force-free field, and then we use this 
  as an initial condition for the second high-resolution grid. Consequently 
  by repeating these procedures, the high-resolution force-free state can be 
  obtained in a short time.
  
  The simulation domain in the ideal force-free case is set to
  $(0,0,0)< (x,y,z) <(2,2,2)$ defined as non-dimensional values, and this is 
  divided into $64^3$grids, $128^3$ grids, $256^3$grids, or $512^3$ grids, 
  case3M1, case3M2, case1M1 and case1M2 shown in table \ref{tbl-1} are applied
  for the multi-grid-type method, whereas direct calculation is applied for 
  case0-case5, without it. On the other hand, for the MHD solution, 
  the entire numerical domain is set to 
  $(0,0,0)<(x,y,z)<(43.2, 43.2, 32.4)$(Mm$^3$) following 
  \cite{2012ApJ...748...53M}, and extracted from the original data; then the 
  total grid number is assigned as $80\times 80\times 60$. All the parameters 
  in each case are given in table \ref{tbl-1}. All of the physical values are 
  normalized using $L_0 = 43.2$(Mm) and $B_0 = 262$(G)(see 
  \citealt{2003ApJ...586..630M} for details); consequently, the numerical 
  domain is set to $(0,0,0)<(x,y,z)<(1,1,0.75)$ in non-dimensional space.

  \section{Result of the NLFFF Extrapolation of Low $\&$ Lou Solution}
  \subsection{Role in the Cleaning of the Numerical Error Related to 
             $\vec{\nabla}\cdot \vec{B}$}
\begin{figure}
  \epsscale{1.}
  \plotone{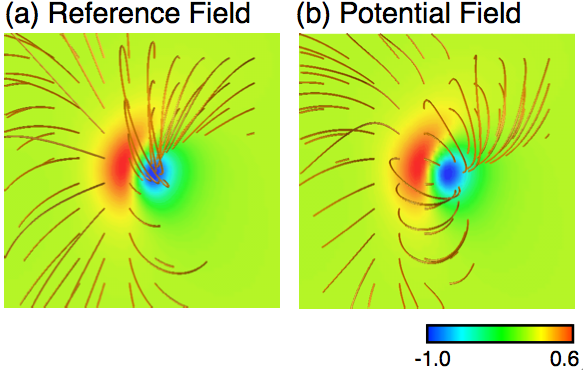}
  \caption{
           (a) Semi-analytical force-free solution introduced by 
               \cite{1990ApJ...352..343L}. Orange lines and color map 
               represent magnetic field lines and distribution of normal 
               component of magnetic field($B_z$), respectively. 
           (b) 3D field lines of potential field extrapolated from normal 
               component of magnetic field on all the boundaries. 
           }
  \label{f1}
  \end{figure}

   We first check the accuracy of the numerical code for the ideal force-free 
  solution introduced by \cite{1990ApJ...352..343L}. A 3D view of this 
  solution is shown in Figure \ref{f1} (a). The lines and background color 
  indicate the magnetic field lines and distribution of the normal component 
  of the magnetic field, respectively. Figure \ref{f1}(b) shows the potential 
  field extrapolated from the normal component of the magnetic field on all 
  the boundaries, which is used as an initial condition in the NLFFF 
  calculation.We calculated the three cases denoted as case $i$ where $i$ = 
  0$-$2. Case0 corresponds to the boundary condition EX, where we also do not 
  use equation(\ref{div_eq}). Case1 and case2 correspond to the boundary 
  conditions Ex and RW, respectively. More detailed informations on case1 
  and case2 are given in table \ref{tbl-1}.

\begin{figure}
  \epsscale{1.}
  \plotone{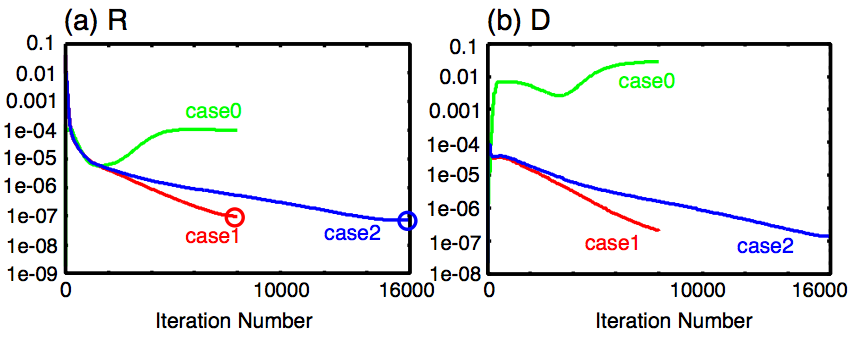}
  \caption{
          (a) Iteration profiles of 
              R = $\int |\vec{J} \times \vec{B}|^{2}dV$ 
              for case0, case1, and case2 are plotted in green, red, and 
              blue lines, respectively. Case0 and case1 use the EX type 
              boundary condition, but case0 does not employ  equation 
              (\ref{div_eq}). Case2 uses the RW boundary condition. Red and 
              blue circles indicate minimum values in case1 and case2.
          (b) Iteration profiles for 
              D= $\int|\vec{\nabla} \cdot \vec{B}|^{2}dV$ for 
              each case with same format as in (a).
          }
  \label{f2}
  \end{figure}

   Figures \ref{f2} (a) and (b) show iteration profiles of  
  $R=\int|\vec{J}\times\vec{B}|^2dV$ and 
  $D=\int|\vec{\nabla}\cdot\vec{B}|^2dV$ 
  for different cases. We clearly found that case1 and case2  tends toward a 
  force-free state, because the R and D profiles decrease with each iteration.
  Iteration was stopped when R reached a minimum value. Case0 shows a much 
  different profile from those of case1 and case2. This result indicates that 
  an iteration profile approaching a force-free state is very sensitive to 
  numerical errors in the deviation from $\vec{\nabla}\cdot \vec{B}=0$. The 
  difference between case1 and case2 is determined by the differences in the 
  lateral and top boundary conditions between them. Even though incomplete 
  lateral and top boundaries in RW are given in case2, the values of R and D 
  are found to be equal order of magnitude of that in case1. However, case2 
  takes about twice as long as case1 to search for the force-free solution.
 
  \subsection{Topology Analysis of the 3D Magnetic Field Lines}
\begin{figure}
  \epsscale{.75}
  \plotone{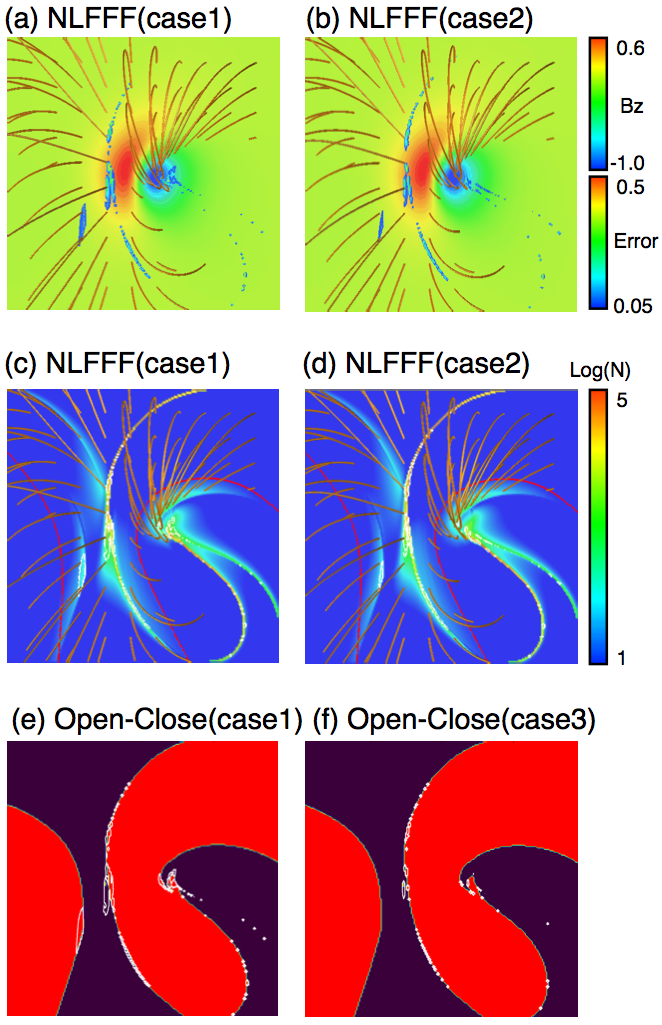}
  \caption{
           (a)-(b)  3D field lines of NLFFFs in case1 and case2. This format 
                    is the same as in Figure \ref{f1} except that the color 
                    contour represents the connectivity error defined in 
                    equation (\ref{error}). 
           (c)-(d)  Field lines and connectivity errors (white contours with 
                    0.05) plotted over the distribution of log(N), where N is 
                    defined in equation (\ref{qsl}).
           (e)-(f)  Connectivity errors in the same format as in (c) or (d) 
                    plotted over a map of the open-closed field lines in 
                    cases 1 and 3, respectively. Closed field line lie in red 
                    regions; other regions are occupied by open field lines.
          } 
    \label{f3}
    \end{figure}

  The three-dimensional NLFFF structures for case1 and case2 are shown 
  in Figures \ref{f3}(a) and (b), respectively. The Color contours 
  represent a connectivity error that is defined as
  \begin{equation}
  \Delta = |\Delta_{Exact} - \Delta_{NLFFF}|.
  \label{error}
  \end{equation}
  $\Delta_{Exact}$ ($\Delta_{NLFFF}$) is the distance from one magnetic field 
  line footpoint to another measured on the bottom surface in the exact(NLFFF)
  solution. The NLFFF solutions in case1 and case2 seem to have almost the 
  same configuration as that of the exact Low $\&$ Lou solution shown in 
  Figure \ref{f1}(a). The connectivity errors between these cases also have 
  the same distributions, a random distribution in the entire domain.  
     
   We investigate the magnetic topology to clarify the cause of the 
  connectivity error. We used the photospheric cross-section of the 
  quasi-separatrix layers (QSLs) introduced by \cite{1996A&A...308..643D}. 
  We calculated the following quantity at each pixel on the vector field maps: 
  \begin{equation}
  \displaystyle N(x,y) = \sqrt{ 
                         \sum_{i=1,2}
                         \left[ 
                           \left( \frac{\partial X_i}{\partial x} \right)^2 
                       +   \left( \frac{\partial X_i}{\partial y} \right)^2 
                         \right]
                              }, 
  \label{qsl}
  \end{equation} 
  where $(X_1, X_2)$ is the relative distance corresponding to 
  $(x'-x'', y'-y'')$. $(x',y')$ and $(x'',y'')$ are the positions of the end 
  points of the field lines whose starting points are two adjacent grid points
  located at ($x'_0$, $y'_0$) and ($x''_0$, $y''_0$) on the photospheric 
  surface. This means that the locations of the end points of these field 
  lines, which are traced from these start points across a large N(x,y) value, 
  may differ greatly.
  
  Figures \ref{f3}(c) and (d) show the connectivity error in  white contours 
   whose magnitude corresponding to 0.05 over the distribution log(N) 
  mapped on the bottom surface. We clearly see that the connectivity errors 
  are almost on the enhancement layers at considerable distance from a 
  polarity inversion line. From this analysis, we found that, remarkably, the 
  error in the NLFFF appears in particular regions where the magnetic topology
  is changing dramatically. 
  
  We show these particular regions in detail. Figures \ref{f3}(e) and (f) show 
  the connectivity errors in the same format as Figures \ref{f3}(c) and (d)
  over a map of the open-closed field lines in  case1 and case3, respectively.
  Closed means that both footpoints of each field line are anchored in the 
  bottom surface; for the open field, one footpoint goes through the side or 
  top boundaries. These results clearly show that the connectivity errors 
  appear along the boundaries between open and closed field lines. On the 
  other hand, the values obtained from this study are $\sim 0.5$, but most of 
  regions are occupied by values less than 0.25, which is much smaller than 
  the entire length of the numerical domain. Consequently, this is not due to 
  a change in the topology from open to closed lines or vice versa; rather, 
  each outer loop of the open or closed field lines deviates slightly from 
  the reference field. Furthermore, Figure \ref{f3}(f) shows a plot in the 
  same format with a higher resolution than that of Figure \ref{f3}(e), 
  which can reduce the error distribution.

  \subsection{Quantitative Comparison of Low $\&$ Lou solution and NLFFF}
\begin{figure}
  \epsscale{1.05}
  \plotone{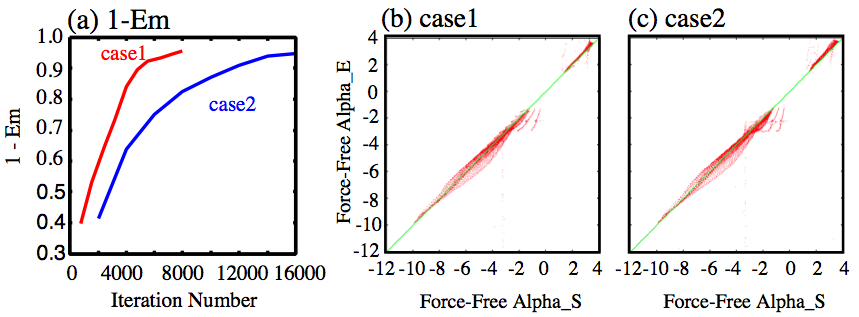}
  \caption{
          (a) Iteration profiles of 1-Em for case1 and case2(red and blue 
              lines, respectively), which are defined by equation 
              (\ref{eq_qtv}). 
          (b) Distribution of force-free $\alpha$ in case1. Horizontal and 
              vertical axes represent the values of the force-free $\alpha$ 
              at each footpoint of each field line. If this distribution is 
              completely along the green line ($y=x$), the magnetic field 
              completely satisfies a force-free state. 
          (c) Distribution for case2 with same format as in (b).
          }
  \label{f4}
  \end{figure}

   We further performed a detailed quantitative analysis as introduced 
   by \cite{2006SoPh..235..161S}. When $\vec{B}$ and $\vec{b}$ represent 
  the semi-analytical Low $\&$ Lou solution and the extrapolated solution, 
  respectively, the accuracy of the NLFFF is estimated by the following 
  sequential relations; 
  \begin{equation}
   C_{vec} \equiv \frac{\sum_{i} \vec{B}_i\cdot \vec{b}_i}
           {\left( \sum_{i}|\vec{B}_i|^2 \sum_{i}|\vec{b}_i|^2 \right)^{1/2}},
  \end{equation} 

  \begin{equation}
   C_{cs} \equiv \frac{1}{N} \sum_i\frac{\vec{B}_i \cdot \vec{b}_i}
      {|\vec{B}_i||\vec{b}_i|},
  \end{equation}

\begin{equation}
   1- E_N \equiv  1-\frac{\sum_i |\vec{b}_i - \vec{B}_i|}{\sum_i |\vec{B}_i|},
  \label{eq_qtv}
  \end{equation}

  \begin{equation}
   1 - E_M \equiv 1-\frac{1}{N}\sum_i \frac{ |\vec{b}_i - \vec{B}_i|}{|\vec{B}_i|},
  \end{equation}

  \begin{equation}
  \epsilon \equiv \frac{\sum_i |\vec{b}_i|^2}{\sum_i|\vec{B}_i|^2},
  \end{equation}
   where $C_{vec}$ is the vector correlation, $C_{cs}$ is the Cauchy-Schwarz 
   inequality, $E_M$ is the mean vector error, $E_N$ is the normalized vector 
   error, $\epsilon$ is the energy ratio, and N is the number of vectors in 
   the field. These results are summarized in table.\ref{tbl-2}. 
   Figure \ref{f4}(a) shows the iteration profiles of 1$-E_m$ for case1 and 
   case2. The final values reach 0.95 in both cases. We clearly see that both 
   case1 and case2 can reconstruct the original Low $\&$ Lou solution with 
   good accuracy and no significant difference is found between them even 
   though case2 takes a longer calculation time than case1.
    
    Finally, we performed another quantitative analysis by evaluating the 
   force-free $\alpha$ in both footpoints of each field line. Because the 
   value of the force-free $\alpha$ should be constant along the field line 
   (cf., equation (\ref{ff_eq_2})) their values at the both footpoints of each 
   field line should be equal in order to satisfy the force-free condition.  
   The force-free $\alpha$ in both footpoints for case1 and case2, measured 
   on the surface above first grid above the bottom one, are mapped in 
   Figures \ref{f4}(b) and (c), respectively. The horizontal and vertical 
   axes represent the values of the force-free $\alpha$ at each footpoint 
   where the diagonal green line corresponds to $y=x$. If an extrapolated 
   field completely satisfies the force-free state, the force-free $\alpha$ 
   will be distributed along this line. Because most points in case1 and 
   case2 are along the green line, this result clearly shows that these cases 
   almost satisfy the force-free state well.  

  \subsection{Multi-Grid Strategy}
  \subsubsection{Procedure of the Multi-grid type method}
   We present a procedure for a multi-grid-type method, which is needed to 
   accelerate the calculation time for high-resolution magnetogram data 
   obtained from {\it e.g.,} SOT/{\it Hinode}. Some algorithms have already 
   implemented it, and an accelerated calculation speed was reported   
   ({\it e.g.,} \citealt{2008SoPh..247..269M} and 
   \citealt{2012ApJ...749..135J}). First, we extrapolate an NLFFF with the 
   coarsest grid $n_{min}^3$, to rapidly propagate large-scale information 
   from a boundary into an interior domain. When the value of R 
   ($\int |\vec{J}\times \vec{B}|^2dV$) reaches a minimum the grid  number is 
   changed to $(2n_{min})^3$ by using a linear interpolation; the exact 
   boundary conditions are maintained, and its location and other parameters 
   are fixed as those in the previous calculation except that $\eta_0$ set to 
   zero. This process is repeated until a force-free field is achieved under 
   given grid points with the highest resolution; therefore, this method is 
   not a full  multi-grid method. The detailed information is given in the 
   Table.\ref{tbl-1}.

  \subsubsection{Accuracy and Calculation Time in Run1}
   We performed our calculations using two different patterns, {\it i.e.,} 
  run1 and run2, using a multi-grid-type method whose final results are 
  obtained in three steps. First, we examined the performance related to run1 
  with an initially assignment of the coarsest grid number $128^3$ grid points
  denoted as case3. After attaining a force-free state with this number of 
  grid points, a higher resolution of ($256^3$) was obtained by using the 
  force-free field realized in the previous step as an initial condition, 
  which is referred to case3M1. In the same way, the highest resolution of 
  ($512^3$) was achieved in case3M2. For comparison with case3M1, we also 
  calculated case4, in which $256^3$ grid points are assigned, but without 
  an implementation of the multi-grid-type method.    

\begin{figure}
  \epsscale{1.}
  \plotone{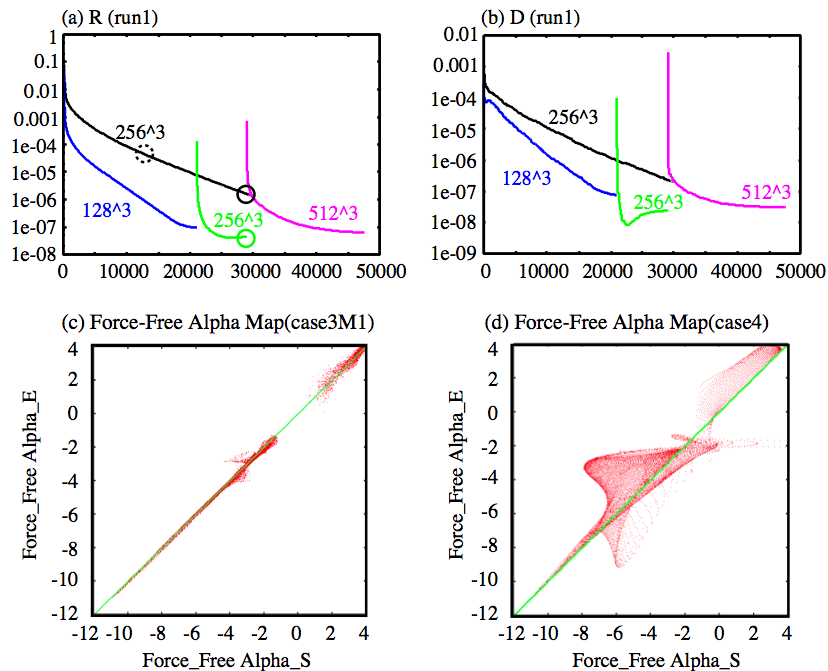}
  \caption{
           (a) Iteration profiles for 
               $R=\int |\vec{J} \times \vec{B}|^{2}dV$ in run1 using 
               $128^3$ (case3), $256^3$ (case3M1), and $512^3$ (case3M2) grid 
               points, shown in the blue, green, and purple, respectively. 
               Black line shows the result for case4, in which the 
               multi-grid-type method is not applied and the number of grid
               points is fixed at $256^3$. Green and black solid circles show 
               the end of the calculation using $256^3$ grid points in case3M1
               and case4, respectively; black dotted circle indicates the 
               result of 12,500 iterations in case4, corresponding to the same
               calculation time with 3.0$ \times 10^{4}$ iterations in 
               case3M1(green solid circle).        
           (b) Iteration profiles for  
               $D= \int |\vec{\nabla} \cdot \vec{B}|^{2}dV$ in run1 in same 
               format in (a). 
           (c) Distribution of force-free $\alpha$ obtained from case3M1, 
               whose state is marked by the green circle in (a). Format is the 
               same as in Figure \ref{f4}.
           (d) Distribution of force-free $\alpha$ obtained for case4 after
               12,500 iterations, marked by black dotted circle in (a).
           }
  \label{f5}
  \end{figure}

  Figures \ref{f5}(a) and (b) show the results related to iteration profiles 
  of 
  $R= \int|\vec{J} \times \vec{B}|^2dV $ and 
  $D= \int|\vec{\nabla}\cdot\vec{B}|^2dV$
  corresponding to run1. The coarsest grid level $128^3$ gradually decreases 
  by $1.0 \times 10^{-7}$ at about $2.0 \times 10^{4}$ iterations; 
  the calculation takes 12h\footnote{Numerical code was parallelized 
  by Message Passing Interface (MPI) and the calculation speed was measured 
  by using a 3.06GHz Xeon X5500 eight-core processor implemented in DELL 
  T7500.} in real time and $1-E_m$ reaches about 0.98, as shown in 
  Table.\ref{tbl-2}. Although R and D suddenly increase as the grid number 
  changes from $128^3$ to $256^3$, they immediately decrease by about 
  $1.0 \times 10^{-7}$ again. This sudden increment in R and D due to the 
  change in the grid is clearly the result of a numerical error arising from 
  an interpolation. However, this error rapidly decreases within  $10^{3}$ 
  iterations, as this scale is small compared to the previous grid, so the 
  diffusion may be effective in decreasing the numerical error associated 
  with a higher mode. The green line corresponding to $256^3$ grids can reach 
  $1.0 \times 10^{-7}$ at about $3.0 \times 10^{4}$ iterations, which is 
  marked by the green circle; the total calculation time takes about 
  45 h in  real time. On the other hand, in case4, which is also assigned to 
  the $256^3$ grids but without the multi-grid-type method, the value of 
  R marked after a total calculation time of 100 h, by a black circle, is 
  found to be one order larger than that for case3M1. The multi-grid-type 
  method significantly reduces the calculation time. Hence, we clearly see 
  that it is an effective method for analyzing high resolution data. The final
  state, plotted by the purple lines, can achieve a high-resolution force-free
  field given by $512^3$ grids.  
  
   Figures \ref{f5}(c) shows a distribution map of the force-free $\alpha$ for
  case3M1 at the green solid circle in Figure \ref{f5}(a); the map is in the 
  same format as Figures \ref{f4}(b) or (c). We see that many red points appear
  along the green lines although a slight deviation appears in the range of 
  0$< \alpha <$2.0. However, the overall pattern of the extrapolated field 
  satisfies a state close to the force-free state. On the other hand, 
  Figure \ref{f5}(d) shows the results for the case4, marked by the dotted 
  black circle in Figure \ref{f5}(a), where the calculation time is the same 
  as that for $3.0\times 10^{4}$ iterations at the end of the calculation in 
  case3M1. As expected, many points deviate from the force-free state. Thus, 
  an implementation of the multi-grid-type method yields a force-free state 
  in a dramatically short time.
  
  \subsubsection{Accuracy and Calculation Time in Run2}
  The procedure for run2 is basically the same as that of run1 except for the 
  assigned grid numbers. In run2 $64^3$ grids points are initially assigned,  
  corresponding to the coarsest grid and initial condition in case1. 
  Eventually, following the same procedure in run1, we obtain a force-free 
  state with $128^3$ and $256^3$ grid points, which are called case1M1 and 
  case1M2, respectively.   

\begin{figure}
  \epsscale{1.}
  \plotone{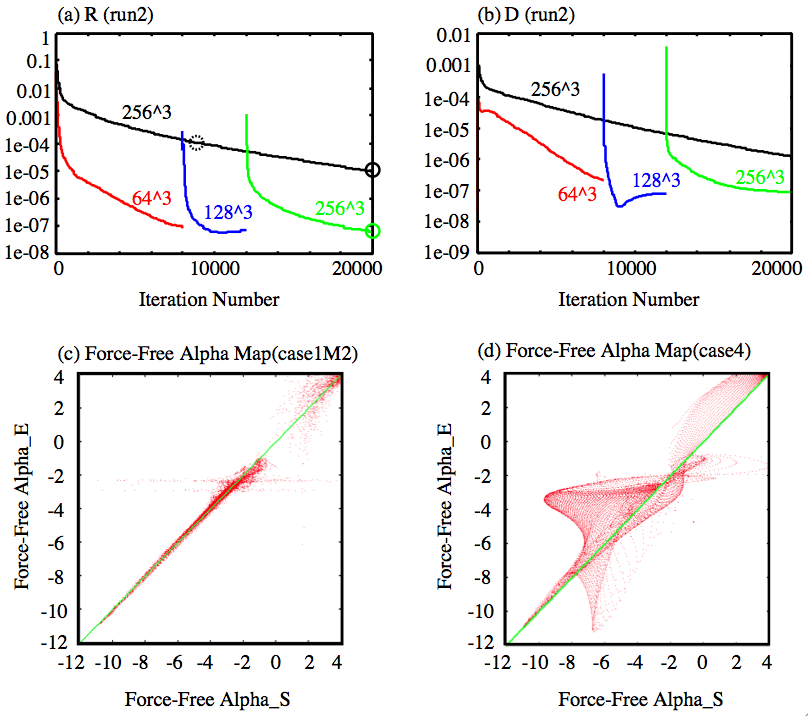}
  \caption
         {
          (a) Iteration profiles of 
              $R=\int |\vec{J} \times \vec{B}|^{2}dV$ 
              in run2; format is essentially the same as in 
              Figure.\ref{f5}(a). Initial grid number differs from that of 
              run1 changing from $64^3$ (case1, red) to $256^3$ 
              (case1M2, green) through $128^3$ (case1M1, blue). Black line 
              shows result for case4, which is the same as in 
              Figure \ref{f5}(a). Green and black solid circles represent 
              2.0$ \times 10^{4}$ iterations for case1M2 and case4, 
              corresponding to the end of calculation with $256^{3}$ grid 
              points. Black dotted circle represents 8625 iterations, 
              corresponding to the same calculation time as at the end of 
              case1M2.  
          (b) Iteration profile of 
              $D= \int |\vec{\nabla} \cdot \vec{B}|^{2}dV$ in run2.  
              (c)-(d) Distributions of force-free $\alpha$ obtained 
              from case1M2 and case4, marked by the green solid and black 
              dotted circles in (a), respectively. These states are obtained 
              in the same calculation time in each case.
         }
  \label{f6}
  \end{figure}

  Figures \ref{f6}(a) and (b) show the results on the R and D profiles, 
  respectively, for each case (case1, case1M1, and case1M2). The black line 
  corresponds to case4, which is the same as in run1. The green and black 
  solid circles represent 2.0$ \times 10^{4}$ iterations, corresponding to 
  the end of the calculation with $256^{3}$ grid points, for case1M2 and 
  case4, respectively. The total calculation times were approximately 
  30 h and 67h, respectively. Further, the black dotted circle represents 
  8625 iterations for case4, corresponding to the same calculation time as 
  that required for 2.0$ \times 10^{4}$  iterations of the multi-grid-type 
  case.  Although the R and D profiles in multi-grid cases reach to values 
  of less than $1.0 \times 10^{-7}$, as with the previous multi-grid-type 
  case, run1, the quantitative value $1-E_m$ shown in table \ref{tbl-2} does 
  not increase from its initial value of 0.95 (obtained from the initial 
  coarsest grid) with increasing grid numbers. We further checked the 
  distribution map of the force-free $\alpha$.

  Figures \ref{f6}(c) and (d) show a distribution map of the force-free 
  $\alpha$ for case1M2 and case4, marked by the solid green and dotted black 
  circles, respectively, in Figure \ref{f6}(a);the format is the same as in 
  Figures \ref{f5}(c) or (d). Case1M2 in the region $\alpha < 0$ is found to 
  yield a better reconstruction in a short time than case4. However, for 0 
  $< \alpha <$ 3.0, this case, as well as case3M1, seems to deviate slightly 
  from the force-free state. In comparison, Figure \ref{f4}(b), which shows 
  a distribution map of the force-free $\alpha$ in the initial state, shows 
  a better force-free state than case1M2 even for $\alpha > 0$.  Hence, this 
  error is clearly derived from an interpolation through a change in the grid 
  that critically affects the value of $1-E_m$, as reported by 
  \cite{2012ApJ...749..135J}.  
  
  \subsection{2D Distribution of the Force-Free $\alpha$}
\begin{figure}
  \epsscale{1.}
  \plotone{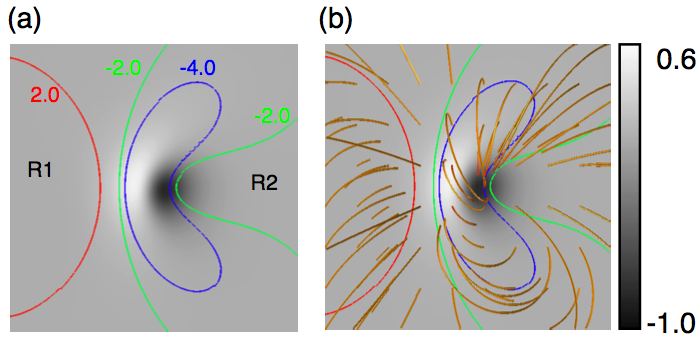}
  \caption
         {
         (a) Selected contours of the force-free $\alpha$ plotted over the 
             $B_z$ component in gray scale. Red, green, and blue lines 
             represent the strengths 2.0, $-$2.0, and $-$4.0, respectively. 
             R1 and R2, enclosed by the red and green 
             lines, fall within 2.0$< \alpha <$ 4.0 and  
             $-$2.0 $< \alpha <$ 0, respectively.
         (b) Magnetic field lines (orange lines) plotted over (a). 
          }
  \label{f7}
  \end{figure}

  Figure \ref{f7}(a) plots contours of the force-free $\alpha$ to clarify why 
  the reconstructed field in the region of weak force-free $\alpha$ 
  (0$<|\alpha|<$3.0) deviated from the force-free state when the 
  multi-grid-type method is used, as such as shown in Figures \ref{f5} and 
  \ref{f6}. The red, green, and blue contours indicate strengths of the 
  force-free $\alpha$ corresponding to 2.0, $-$2.0, and $-$4.0, respectively. 
  From Figure \ref{f7}(a), a region of strong force-free $\alpha$ appears in 
  the central region, where the extrapolated field satisfies the force-free 
  state well, as shown in the previous results. On the other hand regions of 
  weak negative and positive force-free $\alpha$, R1 and R2 lie at 
  considerable distances from the central region in the numerical domain where
  the force-free $\alpha$ is distributed in a range of 2.0$< \alpha <$ 4.0 
  and $-$2.0 $< \alpha <$ 0, respectively. Figure \ref{f7}(b) also shows the 
  magnetic field lines, most of which in  R1 and R2 are rooted in a region 
  near the boundaries of the domain; {\it i.e.,} the numerical errors 
  remarkably appear near the boundaries through an interpolation accompanying 
  a change in the grid number. However, as in Figure \ref{f6}, the 
  extrapolated field in the strong force-free $\alpha$ regions exhibits a 
  better force-free state even when the multi-grid process is used. Thus, the 
  core region in particular can be reconstructed with good accuracy in 
  a dramatically short time by using it.
   
  \section{NLFFF Extrapolation in a Flux Emergence Region Produced by 
           the MHD Simulation}
  We found that our NLFFF extrapolation method performed remarkably well in 
  reproducing an ideal force-free state. Next, we applied it to a  
  flux-emergence region obtained from an MHD simulation 
  (\citealt{2012ApJ...748...53M}). The idea was to check its performance for a
  region that is quite close to the real corona. The simulation
  results of the \cite{2012ApJ...748...53M} provided a hypothetical state of 
  the solar corona affected by the pressure, gravity field, and non-equilibrium
  state that differs greatly from the ideal force-free field introduced by 
  Low $\&$ Lou. In this study, we focus on how well the sheared and twisted 
  field lines in the lower corona are reconstructed by our NLFFF method; most 
  of the free energy is accumulated in these lines, and they are treated as 
  the most important parts for solar active phenomena such as solar flares 
  and CMEs. In the following section, we quantitatively compare the 
  differences in the 3D configurations yielded by the MHD solution and NLFFF 
  extrapolation and finally estimate the degree of twist in them.

  \subsection{Overview of the Active Region from the MHD Simulation}    
  \cite{2012ApJ...748...53M} surveyed the dynamics of flux emergence with  
  respect to a wide range of parameters set at the initial time of the 
  twisted magnetic flux tube. In this study, we select one snapshot at the 
  last moment of the MT case (see Table 1 in their paper), in which a flux 
  tube embedded in the convection zone has emerged into the solar corona and 
  formed coronal magnetic loops. First, we introduce the basic components of 
  the magnetic field obtained from the MHD simulation, which is used as the 
  boundary condition in our NLFFF extrapolation method.
  
\begin{figure}
  \epsscale{1.}
  \plotone{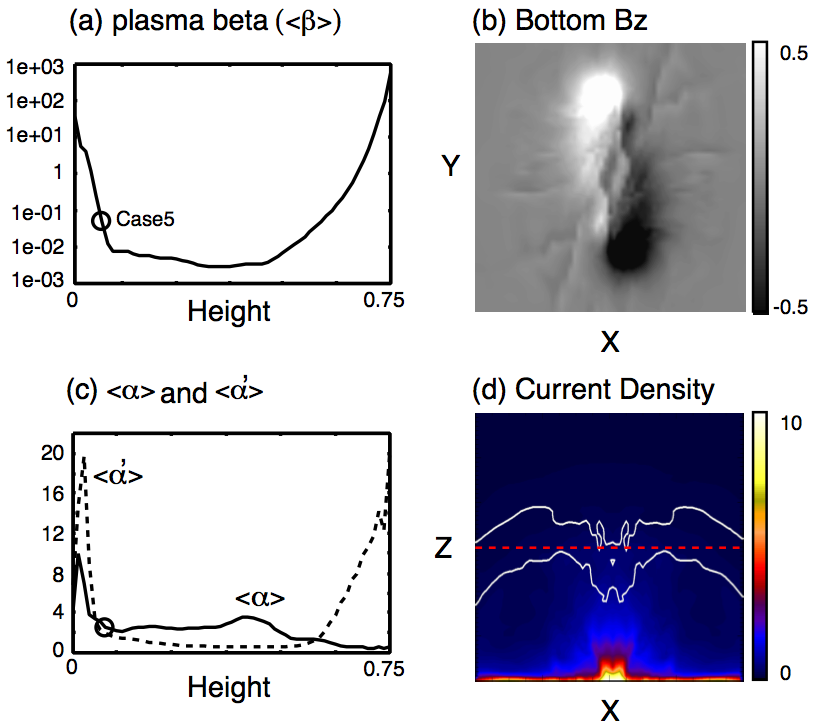}
  \caption
          {
          (a) Height profile of integrated plasma $\beta$($<\beta(z)>$). Black
              circles indicate case5. 
          (b) Distribution of $B_z$ component at 2700 (km) above the 
              photosphere plotted in gray scale, which is obtained from the 
              flux-emergence simulation in the MT case in 
              \cite{2012ApJ...748...53M}. This magnetic field is normalized 
              by the maximum value($B_0$=262(G)); {\it i.e.}, the maximum and 
              minimum values correspond to 1 and $-$1.
          (c) Height profiles of integrated force-free $\alpha$ 
              $<\alpha(z)>$ (solid line) and non-force-free component 
              $<\alpha(z)^{'}>$ (dashed line) according to the equations 
              (\ref{ff_c}) and (\ref{nff_c}).
          (d) Velocity profile ($\int \vec{|v|}dy$) in white contours plotted 
              over distribution of current density ($\int \vec{|J|}dy$). 
              Strength of contours is 8.0. Red dotted line indicates
              half-height of entire domain.  
           }
 \label{f8}
 \end{figure}

   Figure \ref{f8}(a) shows a height profile of the integrated plasma 
  $\beta$ defined as $<\beta(z)>=2<P(z)>/<B^2(z)>$), where $<P(z)>$ and 
  $<B^2(z)>$ represent the average plasma and magnetic pressure, respectively,
  in a horizontal plane; $<P(z)>=\int_{S_{xy}} P(x,y,z) dxdy$ and 
  $<B^2(z)> = \int_{S_{xy}} B^2(x,y,z) dxdy$. Near the bottom surface, a 
  high-$\beta$ ($1 \leq <\beta> \leq 100$) regime is formed, which is suddenly 
  converted into a low-$\beta$ regime ($<<1$) around an intermediate height 
  and finally again reaches a value similar to that at the bottom boundary as 
  the height increases. We select the boundary condition at $<\beta>=0.06$ 
  which corresponds to case5 (marked by solid circles). The distribution 
  of the normal component of the magnetic field in case5 is shown in 
  Figure \ref{f8}(b). It is formed by the emerging flux tube at 2700(km) 
  above photosphere in the MHD simulation. The value of the magnetic field is 
  normalized by its maximum value at this height. Hereafter, this height is 
  regarded as the bottom surface.
  
   Figure \ref{f8}(c) shows height profiles of the integrated force-free 
  $\alpha$ ($<\alpha>$) and the non-force-free component ($<\alpha^{'}>$), 
  denoted as 
  \begin{equation}
  <\alpha> = \int_{S_{xy}} \frac{|\vec{J} \cdot \vec{B}|}{|\vec{B}|^2} dxdy,
  \label{ff_c}
  \end{equation}

  \begin{equation}
  <\alpha^{'}> = \int_{S_{xy}} \frac{|\vec{J} \times \vec{B}|}{|\vec{B}|^2} dxdy.
  \label{nff_c}
  \end{equation}
  The value of $<\alpha>$ is larger than that of  $<\alpha^{'}>$ in those 
  regimes where the condition of low $<\beta>$ is satisfied well. The 
  solid circle indicates the result for the case5, where $<\alpha>$ is 
  competing with $<\alpha'>$.

  However, this MHD solution does not satisfy the equilibrium state 
  completely; therefore, we have to estimate the degree to which the Lorentz 
  force is accumulated in the selected boundary condition, as shown in 
  Figure \ref{f8}(b). We estimate it using the following values: 
  \[
  \epsilon_{force} = \frac{ (|\int_S B_xB_z| + |\int_S B_yB_z| 
                           + |\int_S((B_x^2 + B_y^2)-B_z^2) |)dxdy}
                          {\int_S (B_x^2 + B_y^2 + B_z^2)dxdy},
  \]

  \[
  \epsilon_{torque} = \frac{(|\int_S x((B_x^2 + B_y^2)-B_z^2)| +
                             |\int_S y((B_x^2 + B_y^2)-B_z^2)| +
                             |\int_S yB_xB_z-xB_yB_z|)dxdy}
                           {\int_S \sqrt{x^2 + y^2}(B_x^2 + B_y^2 + B_z^2)dxdy
                           },
  \]
  where S represents a surface on the bottom boundary, and
  $\epsilon_{force}$ and $\epsilon_{torque}$ correspond to the force balance 
  and torque balance parameters, respectively. When $\epsilon_{force}<<1$ and 
  $\epsilon_{torque}<<1$, the boundary surface approximately satisfies the 
  force-free condition (\citealt{2006SoPh..233..215W}). As a result, in this 
  case($\epsilon_{force}=$0.275 and $\epsilon_{torque}=$0.382), they are 
  deviate from the force-free state; nevertheless these values are close to 
  the SP/{\it Hinode} data for Dec.12, 2006 according to 
  \cite{2012SoPh..281...37W}.

  Figure \ref{f8}(d) shows profiles of the integrated velocity with respect 
  to the $y$ direction ($\int |\vec{v}(x,y,z)|dy$) with a map of the current 
  density in the same format as ($\int |\vec{J}(x,y,z)|dy$). We clearly see 
  that the color is strongly enhanced in the lower central area in which the 
  core field is formed. The large velocity fields, plotted by the white 
  contours, are concentrated on the area around the half-height of the entire 
  box, marked by the red dashed line. Consequently, we infer that 
  extrapolation of this region is difficult; however, one of our interests is 
  to address how well the core field is reconstructed under this condition. 
 
  \subsection{Result of the NLFFF}
  \subsubsection{1D Profiles from the NLFFF and MHD Solutions}
\begin{figure}
  \epsscale{1.}
  \plotone{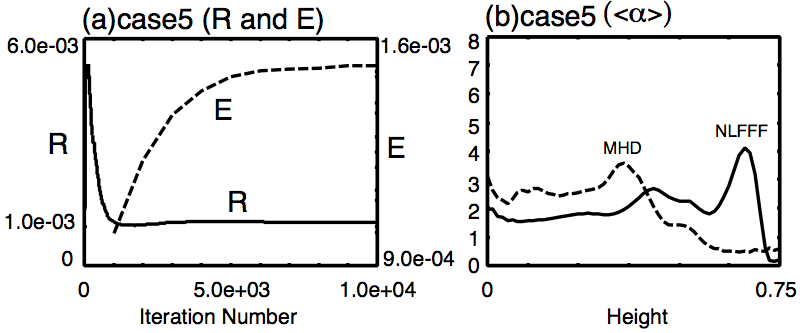}
  \caption
          {
          (a) Iteration profiles of total Lorentz force 
              ($R=\int|\vec{J}\times\vec{B}|^2dV$) and magnetic energy 
              ($E=1/2\int |\vec{B}|^2 dV$) for case5. Solid and 
              dashed lines represent R and E, the values of which are shown 
              on the left and right vertical axes, respectively. 
          (b) Height profile of integrated force-free 
              $\alpha$ ($<\alpha>$) calculated from MHD solution(dashed line)  
              and $<\alpha_{nlfff}>$ from NLFFF(solid line) in case5 
              calculated for (0.25,0.25) $<$ (x,y) $<$ (0,75, 0.75).  
          }
 \label{f9}
 \end{figure}

  First, we show the one-dimensional result from the NLFFF and compare them  
  with the MHD solution. Figure \ref{f9}(a) shows iteration profiles of the 
  total Lorentz force R (solid line) and magnetic energy E (dashed line) 
  corresponding to case5. $\gamma$ as defined in the equation (\ref{step_eq}) 
  is equal to 1, and the other parameters used in this NLFFF calculation are 
  shown in Table \ref{tbl-1}(see case5). The NLFFF is selected at 
  $1.0 \times 10^4$ iterations, at approximately which R and E begin to 
  saturate. 

  Figure \ref{f9}(b) shows the height profiles of $<\alpha>$ in the MHD 
  solution (dashed line) and $<\alpha_{nlfff}>$ in NLFFF (solid line) for 
  case5. The values of $<\alpha_{nlfff}>$ and its pattern deviates slightly  
  from those of $<\alpha>$ corresponding to those regimes where the value of 
  $<\alpha>$ is dominant over $<\alpha^{'}>$, as shown in Figure \ref{f8}(b). 
  On the other hand, these values and profiles of $<\alpha_{nlfff}>$ are found
  to deviate greatly from those of $<\alpha>$ in the upper area, above the 
  half-height of the entire domain.
%
  \subsubsection{3D Magnetic Structures and Topologies in the NLFFF and MHD
                 Simulations}
\begin{figure}
  \epsscale{.8}
  \plotone{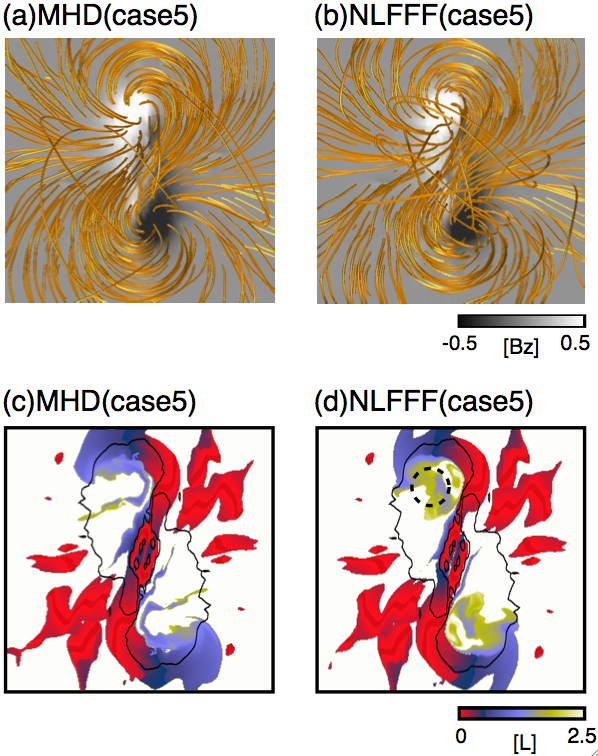}
  \caption
          {
          Selected field lines obtained using (a) MHD solution and (b) NLFFF 
          in case5, plotted over the $B_z$ component. Distribution of field 
          line lengths mapped on the bottom surface obtained from (c) MHD 
          solution and (d)NLFFF. All the field lines are traced from the 
          bottom surface. Colored areas are occupied by closed field lines, 
          both footpoints of which are anchored in the bottom surface, and 
          whose maximum length is $L_{max}=2.5$. White areas are dominated by 
          another type of field lines with one footpoint rooted in the lateral
          boundaries. These are plotted in a range of 
         (0.1,0.1)$<$(x,y)$<$(0.9,0.9). Black solid lines represent a contour 
          for the $|B_z|=0.1$. Dashed circle marks region where closed loops
          are anchored. 
          }
 \label{f10}
 \end{figure}

   Next, we show a 3D view of the MHD simulation and NLFFF and present a
  detailed comparison in terms of the magnetic topology. Figures \ref{f10}(a) 
  shows a top view of the selected field lines obtained from the MHD solutions
  for case5. These are traced from the surface, which is 2700(km) above the 
  photosphere, as shown in Figure \ref{f8}(a). Figure \ref{f10}(b) also shows 
  the selected 3D field lines in the NLFFF, which are also extrapolated from 
  the same surface as those described above. From these results, we infer that
  although they are not exactly the same, the NLFFF seems to reproduce an 
  inverse S shaped structure lying above the polarity inversion line and is 
  qualitatively similar to the MHD solution. It is important to capture these 
  S or inverse S shaped structures because they are considered to be 
  precursors for huge flares (\citealt{1999GeoRL..26..627C}) observed in 
  solar active regions. We further investigate the magnetic topology in more 
  detail to clarify the differences in between the NLFFF and MHD solutions.

  Figures \ref{f10}(c) and (d) show the distribution of the field line length 
  mapped on the bottom surface for the MHD solution and the NLFFF, 
  respectively. One footpoint of the field lines rooted in the white areas is
  not at the bottom surface; that is, their other footpoints is rooted in the 
  lateral boundary surfaces, whereas the other field lines traced from colored
  areas are closed. Therefore, the boundaries between the colored and white 
  areas represent the separatrix separating the closed and open field lines, 
  in which the QSL values are enhanced. The inverse S shaped structure can be 
  formed by the NLFFF as well as by the MHD solution, and this structure is 
  better captured in the NLFFF except in the regions marked by the dashed 
  circle in which the closed loops are anchored. 

   To provide  more clarification, we also show the field line structure 
  within the dashed circle in Figure \ref{f10}(d) in more detail. 
  Figures \ref{f11}(a) and (b) show the field lines (red) in the MHD and 
  NLFFF, respectively, traced from the area marked by the dashed circle in 
  Figure \ref{f10}(d). The field line profiles in MHD and NLFFF are  
  remarkably different. One footpoint of the field line in the MHD solution 
  touches the lateral surface: i.e., the line crosses the boundary surface, 
  whereas both footpoints in the NLFFF touch the bottom surface. 
  Figures \ref{f11}(c) and (d) show the distribution of the height of one 
  footpoint of the field line measured from bottom surface in the MHD and 
  NLFFF. All of the field lines are traced from the bottom surface, and most
  of their footpoints appear in the lower areas plotted in red. On the other 
  hand, we see strong enhancement areas, which are marked by dashed circles, 
  in the MHD solution, whereas these regions are not seen in the NLFFF. These 
  enhanced areas indicate that the height of one footpoint of the field lines 
  is above the half-height of the numerical domain; therefore, it might be 
  difficult for the NLFFF to capture these field lines in this case.
              
  \subsubsection{Magnetic Twist in the NLFFF and MHD Simulations}
   Finally we compare the magnetic twist obtained using the NLFFF with that in 
  the MHD solution. This value represents the degree of twist of a magnetic 
  field line as determined by the measurement of the magnetic helicity 
  generated due to the current parallel to a field line
  (
  \citealt{1984JFM...147..133B}; 
  \citealt{2006JPhA...39.8321B};
  \citealt{2010A&A...516A..49T}; 
  \citealt{2011ApJ...738..161I};
  \citealt{2012ApJ...760...17I};
  \citealt{2013arXiv1304.8073I}  
   ). 
  Because a large amount of magnetic twist can lead to an unstable 
  condition
  (\citealt{1958PhFl....1..265K};
   \citealt{1979SoPh...64..303H};
   \citealt{2004A&A...413L..27T}; 
   \citealt{2005ApJ...630L..97T};
   \citealt{2005ApJ...630..543F}; 
   \citealt{2006ApJ...645..742I};
   \citealt{2006ApJ...645..732B}),
  an estimation of the magnetic twist is important for analyzing the stability
  of the solar coronal magnetic field. We are interested in addressing the 
  extent to which the magnetic twist can be reconstructed. The magnetic twist 
  is defined as,
  \begin{equation}
  T_n = \frac{1}{4\pi} \int \alpha dl, 
  \label{eq_twist}
  \end{equation} 
  where the line integral $\int dl$ is taken along a magnetic field line, and 
  the force-free $\alpha$ is calculated from 
  $\alpha = \vec{J} \cdot \vec{B}/|\vec{B}|^2$. 

   Figure \ref{f12} shows the distributions of the magnetic twist in each 
  field line from the MHD solution, where the NLFFF is also mapped on the 
  surfaces at the same height. Positive and negative values indicate 
  right-handed and left-handed twists, respectively, depending on the value 
  of the magnetic helicity accumulated in an initial flux tube embedded in 
  the subsurface. The black contours show the normal component of the magnetic
  field. We clearly see that the strongly twisted regions in the MHD solution 
  are localized at positive and negative polarities, in which the strong twist
  values over one turn ($|T_n|>1.0$) are found near the vicinity of the tip. 
  In contrast, although the strongly twisted regions in the NLFFF are also 
  localized in the same areas at the both polarities, their distributions and 
  values are not the same as those of the MHD solution. The areas marked by 
  the red circles are remarkably different in both cases; in the NLFFF, closed
  field lines exist in these areas, whereas that type of field line is 
  not seen in the MHD solution. Although the reconstructed twisted lines in 
  other areas tend to capture those in the MHD solution, their values are 
  relatively weak in most regions near the tip at both polarities, whereas 
  a high twist value, $|T_n|>1.0$, is also observed in some areas. 
  These results show that the shape of the sheared field lines is 
  reconstructed well qualitatively; on the other hand, the value of the 
  magnetic twist tends to be weaker than that of the reference field.

  \section{Summary $\&$ Discussion}
   In this study, we developed an NLFFF extrapolation code based on the 
  MHD relaxation method and applied it to the ideal force-free field 
  introduced by \cite {1990ApJ...352..343L}. Our NLFFF extrapolation code 
  generally produces the original ideal force-free state, even though 
  incomplete lateral and top boundary conditions are imposed. Moreover, 
  although errors related to the connectivity of the field lines clearly 
  appeared along the separatrix layers (QSL), their topology is not changed 
  dramatically. We further implemented  the multi-grid-type method in our 
  NLFFF extrapolation code, which increased calculation speed toward a 
  force-free state compared to the code without it. We see that the inner 
  region of the numerical domain in particular can be reconstructed with high 
  accuracy. Thus, our code can be used as a possible method for extrapolating 
  the NLFFF in a shorter time using a high-resolution vector field obtained 
  from SOT/{\it Hinode} or HMI/{\it SDO}.
    
   Next, we also applied our extrapolation method to the MHD solutions 
  obtained from \cite{2012ApJ...748...53M}, which are influenced by the gas 
  pressure, gravity, and non-equilibrium state. In principle, it does not 
  seem appropriate to apply the NLFFF to MHD solutions; nevertheless, in 
  the interest of readers, we also checked the effectiveness of the NLFFF 
  extrapolation for a more realistic situation of the solar corona. As 
  a result, we found that the S-shaped sheared structure formed in the lower 
  corona can be captured well by our NLFFF despite the difference in the value 
  of the twist and the profile compared to the original MHD solutions. Thus, 
  we conclude that our NLFFF extrapolation can work  in the lower coronal 
  region, in which a strong current is stored. 

\begin{figure}
  \epsscale{.85}
  \plotone{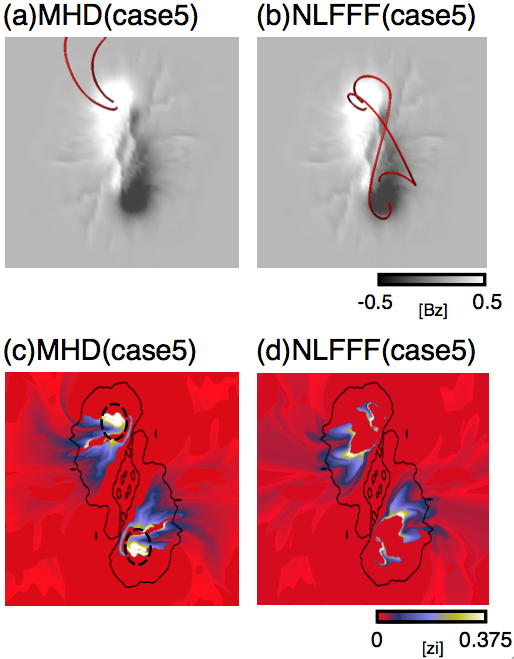}
  \caption{
           Selected field lines obtained using (a) MHD solution and (b) NLFFF 
           for case5 traced from the dashed circle in Figure \ref{f10}(d). The 
           $B_z$ component is plotted in gray scale in the  same format 
           as in Figure \ref{f8}(b). Distribution maps of height of one 
           footpoint in field line measured from bottom surface for (c) MHD 
           solution and (d) NLFFF. Solid lines represent contours for 
           $|B_z|=0.1$. 
          }
  \label{f11}
  \end{figure}

\begin{figure}
  \epsscale{.85}
  \plotone{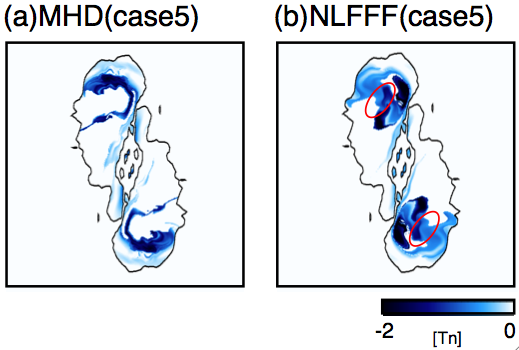}
  \caption
         {
          Twist value of each field line in (a) MHD solution and (b) NLFFF 
          for case5 mapped on the bottom surface. $B_z$ contours are plotted 
          by  solid lines ($|B_z|=0.1$) in the same format as in 
          Figures \ref{f10} or \ref{f11}. The domain is set in the range of 
          (0.2,0.2)$<$(x,y)$<$(0.8,0.8).  
          }
  \label{f12}
  \end{figure}

   Our NLFFF extrapolation may be less effective for reproducing the upper 
  region of the entire numerical domain. Note, however, that the numerical 
  solutions of \cite{2012ApJ...748...53M} describe an extreme situation 
  compared to the real solar corona. For example, a pre-existing coronal field 
  is not assumed in this MHD simulation, and in such a situation the magnetic 
  loops fulling the upper area can expand continuously in all directions, as 
  pointed out by \cite{2004ApJ...605..480M} and \cite{2011ApJ...731..122M}
  (see Figure 8). Therefore, it might be difficult for the NLFFF to 
  reconstruct this type of field line, as shown in our results 
  (Figures \ref{f11} and \ref{f12}). We believe that a consideration of the 
  pre-existing coronal magnetic field may be important in suppressing the 
  expanded loops during the process of flux emergence and achieving a steady 
  state under which an S or inverse S-shaped field may be formed. We expect 
  that our NLFFF extrapolation might be able to reproduce these regions with 
  better accuracy than that in the results presented here. These results are 
  derived from one of the results of \cite{2012ApJ...748...53M}, and our 
  interests are in further addressing the extrapolation of the coronal 
  magnetic field using boundary conditions formed by stronger or weaker 
  twisted flux tubes, or with various plasma $\beta$ values. However, these
  remain as future works.

  On the other hand, a non-force-free extrapolation method has been 
  developed recently by other authors
  ({\it e.g.},\citealt{2006A&A...457.1053W};
              \citealt{2008SoPh..247...87H};
              \citealt{2013ApJ...768..119Z};
              \citealt{2013SoPh..282..283G}).
  These methods are still being improved and are expected to be a solid tool 
  for capturing the MHD solution more accurately in the near future.

   From this study, we conclude that the extrapolated field can robustly 
  reproduce an ideal force-free state, {\it e.g.,} the Low $\&$ Lou solution. 
  In contrast for the MHD solutions obtained from the flux-emergence 
  simulation, this method captures the sheared field, such as elbow-like 
  structures lying in the lower corona, in which the strongest energy is 
  accumulated. Therefore, the extrapolated field may provide a better 
  understanding of active phenomena in the solar corona. Because the 
  SP/{\it Hinode} and HMI/{\it SDO} can observe the vector field with high 
  spatial and temporal resolution,  we will consider a comprehensive view of 
  the flare dynamics and onset mechanism by using the 3D extrapolated magnetic
  field obtained from this data set in a future work.
      
  \acknowledgments
   We extend special thanks to Drs.\ T.\ Miyoshi, and K.\ Hayashi for many 
  helpful discussions and anonymous referee for constructive comments. 
  S.\ I.\ was supported by the International Scholarship of Kyung Hee University. 
  G.\ S.\ C.\ was supported by the National Research Foundation of Korea 
  (NRF-2010-0025403), and he thanks kyung Hee University for granting him 
  a sabbatical in the 2012-2013 school year, during which his part of this 
  research was performed. This research was supported by the BK21 plus 
  program through the National Research Foundation (NRF) funded by the 
  Ministry of Education of Korea, and by the Meteorological Administration of 
  Korea through the National Meteorological Satellite Center. The computational 
  work was carried out within the computational joint research program at the 
  Solar-Terrestrial Environment Laboratory, Nagoya University. Part of computer 
  simulation was performed on the Fujitsu PRIMERGY CX250 system of the 
  Information Technology Center, Nagoya University. Part of computations, 
  data analysis and visualization are performed  using resource of the 
  OneSpaceNet in the NICT Science Cloud.
 



\clearpage

  \clearpage

  \clearpage

    \clearpage

  \clearpage

  \clearpage

  \clearpage
    
  \clearpage

 \clearpage

 \clearpage

 \clearpage

  \clearpage

  \clearpage

  \begin{table}
  \begin{center}
  \caption{These are parameters in all cases.\label{tbl-1}}
  \begin{tabular}{crrrrrrr}
  \tableline\tableline
   Parameters & BC & $R_{min}$ & $v_{max}$ &$c_h$ & $\eta_0$ & d$\gamma$ &
   GN \\
  \tableline
   case1     &  EX  & $1.0\times 10^{-4}$ & 1.0 & 5.0 & $3.75 \times 10^{-5}$  
             &  -   & $64^3$  \\
   case1M1   &  EX  & $1.0\times 10^{-4}$ & 1.0 & 5.0 & 0
             &  -   & $128^3$ \\
   case1M2   &  EX  & $1.0\times 10^{-4}$ & 1.0 & 5.0 & 0
             &  -   & $256^3$ \\
   case2     &  RW  & $1.0\times 10^{-4}$ & 1.0 & 5.0 & $3.75 \times 10^{-5}$
             &  -   & $64^3$  \\
   case3     &  EX  & $1.0\times 10^{-4}$ & 1.0 & 5.0 & $3.75 \times 10^{-5}$
             &  -   & $128^3$ \\ 
   case3M1   &  EX  & $1.0\times 10^{-4}$ & 1.0 & 5.0 & 0 
             &  -   & $256^3$ \\
   case3M2   &  EX  & $1.0\times 10^{-4}$ & 1.0 & 5.0 & 0
             &  -   & $512^3$ \\
   case4     &  EX  & $1.0\times 10^{-4}$ & 1.0 & 5.0 & $3.75 \times 10^{-5}$
             &  -   & $256^3$ \\
   case5     &  RW  & $5.0\times 10^{-3}$ & $5.0 \times 10^{-2}$ & 0.2
             & $5.0 \times 10^{-5}$ & 0.02 & $80\times 80\times 60$ \\
   \tableline
   \end{tabular} 
   \end{center}
   \end{table}

  \begin{table}
  \begin{center}
  \caption{The result of the quantitative analysis, introduced by 
           \cite{2006SoPh..235..161S}, for Low $\&$ Lou solution.
           \label{tbl-2}}
  \begin{tabular}{crrrrrrr}
  \tableline\tableline
  Star & method & $C_{vec}$ & $C_{cs}$ & $1-E_N$ & $1-E_M$ & $\epsilon$ & 
  grid number \\
  \tableline
  Low \& Lou & -               & 1.00 & 1.00 & 1.00 & 1.00 & 1.00 & $64^3$ \\
  Wiegelmann & Optimization    & 1.00 & 1.00 & 0.98 & 0.98 & 1.02 & $64^3$ \\
  case1      & MHD Relaxation  & 1.00 & 1.00 & 0.97 & 0.95 & 1.02 & $64^3$ \\ 
  case2      & MHD Relaxation  & 1.00 & 1.00 & 0.97 & 0.95 & 1.04 & $64^3$ \\
  case3      & MHD Relaxation  & 1.00 & 1.00 & 0.99 & 0.98 & 1.00 & $128^3$\\
  case4      & MHD Relaxation  & 1.00 & 1.00 & 0.98 & 0.96 & 0.99 & $256^3$\\
  case1M1    & MHD Relaxation  & 1.00 & 1.00 & 0.97 & 0.94 & 1.02 & $128^3$\\
  case1M2    & MHD Relaxation  & 1.00 & 1.00 & 0.97 & 0.93 & 1.02 & $256^3$\\
  \tableline
  \end{tabular}
  \end{center}
  \end{table}




\end{document}